\documentclass[aps,prl,twocolumn,superscriptaddress]{revtex4}
\usepackage{graphicx}
\usepackage{amsmath}
\usepackage{cleveref}
\usepackage{color}

\begin{document}

\title{Distinguishing \textcolor{black}{Advective} and Powered Motion in Self-Propelled Colloids}

\author{Young-Moo Byun}
%\email[]{ymbyun@gmail.com}
\affiliation{Department of Physics and Astronomy, University of Missouri, Columbia, MO 65211, USA}
\affiliation{Department of Physics, Pennsylvania State University, University Park, PA 16802, USA}

\author{Paul E. Lammert}
%\email[]{pel1@psu.edu}
\affiliation{Department of Physics, Pennsylvania State University, University Park, PA 16802, USA}

% Update affiliation? (vhc)
\author{Yiying Hong}
%\email[]{yuh118@chem.psu.edu}
\affiliation{Department of Chemistry, Pennsylvania State University, University Park, PA 16802, USA}

\author{Ayusman Sen}
%\email[]{asen@psu.edu}
\affiliation{Department of Chemistry, Pennsylvania State University, University Park, PA 16802, USA}

\author{Vincent H. Crespi}
\email[]{vhc2@psu.edu}
\affiliation{Departments of Physics, Chemistry, and Materials Science and Engineering, Pennsylvania State University, University Park, PA 16802, USA}

\date{\today}

\begin{abstract}
Self-powered motion in catalytic colloidal particles provides a compelling example of active matter, i.e. systems that engage in single-particle and collective behavior far from equilibrium. The long-time, long-distance behavior of such systems is of particular interest, since it connects their individual micro-scale behavior to macro-scale phenomena. In such analyses, it is important to distinguish motion due to subtle \textcolor{black}{advective} effects -- which also has long time scales and length scales -- from long-timescale phenomena that derive from intrinsically powered motion. Here, we develop a methodology to analyze the statistical properties of the translational and rotational motions of powered colloids to distinguish, for example, active chemotaxis from passive \textcolor{black}{advection} by bulk flow.
\end{abstract}

%a bulk flow

\maketitle

Active matter, here taken to mean systems in which the fundamental microscopic degrees of freedom are driven far from equilibrium due to localized free energy input, can display a wealth of behaviors at both the single-particle~\cite{Ramaswamy10, Marchetti13} and collective levels~\cite{Deseigne10, Palacci13}. In such systems, attention naturally turns to long-distance and long-time behaviors, to make connections to both fundamental phenomena such as phase transitions that take place in the thermodynamic limit and potential applications in transport and sensing that may require active motion of analytes, cargo, or motors over long distances. One such long length-scale motion is {\it taxis}, the sustained directed motion of an object up or down a gradient in a scalar field, such as chemical concentration or temperature. The object in question could be a biological system where the chemical gradient is food or poison~\cite{adler69, jeon02, Wadhams04, Roussos11}, or an abiotic self-propelledly propelled nanoparticle~\cite{Baraban13}, microrod~\cite{hong07, sun08, hong10}, or even enzyme~\cite{Dey14} where the gradient may similarly be a chemical fuel or a catalytic promoter/inhibitor. Fluids with solute gradients are obviously non-equilibrium hydrodynamic systems, and as such are prone to \textcolor{black}{advective} effects in addition to any local mobility associated with self-propelledly powered swimmers. Therefore, in the analysis of the long-distance, long-time behavior of active materials it is important to distinguish motion due to subtle \textcolor{black}{advective} effects from phenomena that derive from intrinsically powered motion, such as chemotaxis. Here, we investigate how detailed statistical analysis of swimmer trajectories can distinguish biased active swimming up or down a gradient from passive \textcolor{black}{transport by} bulk flow, even in cases where the requisite directed motions are subtle and much weaker than random Brownian fluctuations.

%~\cite{Salafi17,Baraban13}
%generated by the same gradient

For concreteness, we consider in detail one particular \textcolor{black}{supposed} method of producing such a gradient (used for example in Refs.~\cite{hong07, sun08, hong10}): immersion of a fuel source, such as a hydrogel soaked with H$_2$O$_2$, into an aqueous environment. Hydrogen peroxide is a fuel for bimetallic catalytic microrods that are capable of asymmetric electrochemical decomposition of hydrogen peroxide into water and oxygen across their two metal surfaces~\cite{paxton04, paxton06, wang06}. This system provides one example of a broader class of powered swimmers that drive motion along a structurally defined symmetry axis (here defined by the microrod axis, which is also the axis connecting the two metals). A Pt/Au microrod \textcolor{black}{($\sim$2~$\mu$m long and $\sim$0.37~$\mu$m in diameter)} immersed in an aqueous solution of H$_2$O$_2$ propels itself predominantly along its long axis either due to self-electrophoresis~\cite{wang06} (at small scales) or oxygen bubble recoil~\cite{mirkovic10a} (at much larger scales). As noted above, a localized fuel source has the potential to induce hydrodynamic flow as well. In the particular case of a hydrogel fuel source, the capacity of the hydrogel to absorb water depends on the ionic strength of the ambient solution, with greater capacity in solutions of lower ionic strength~\cite{Kabiri11, Ahmed15}. Both \textcolor{black}{putative} chemical gradients and \textcolor{black}{advective} flows will thus be sensitive to the initial conditions (e.g. initial fuel concentration and distribution, distance of swimmer from fuel source) and will evolve with time as the system approaches equilibrium. Thus, it is important to find robust methods to distinguish these two possible sources of directed motion towards or away from a fuel source, one active and one passive.

\begin{figure}
\begin{tabular}{c}
\includegraphics[trim=0mm 0mm 50mm 0mm, clip, width=0.40\textwidth]{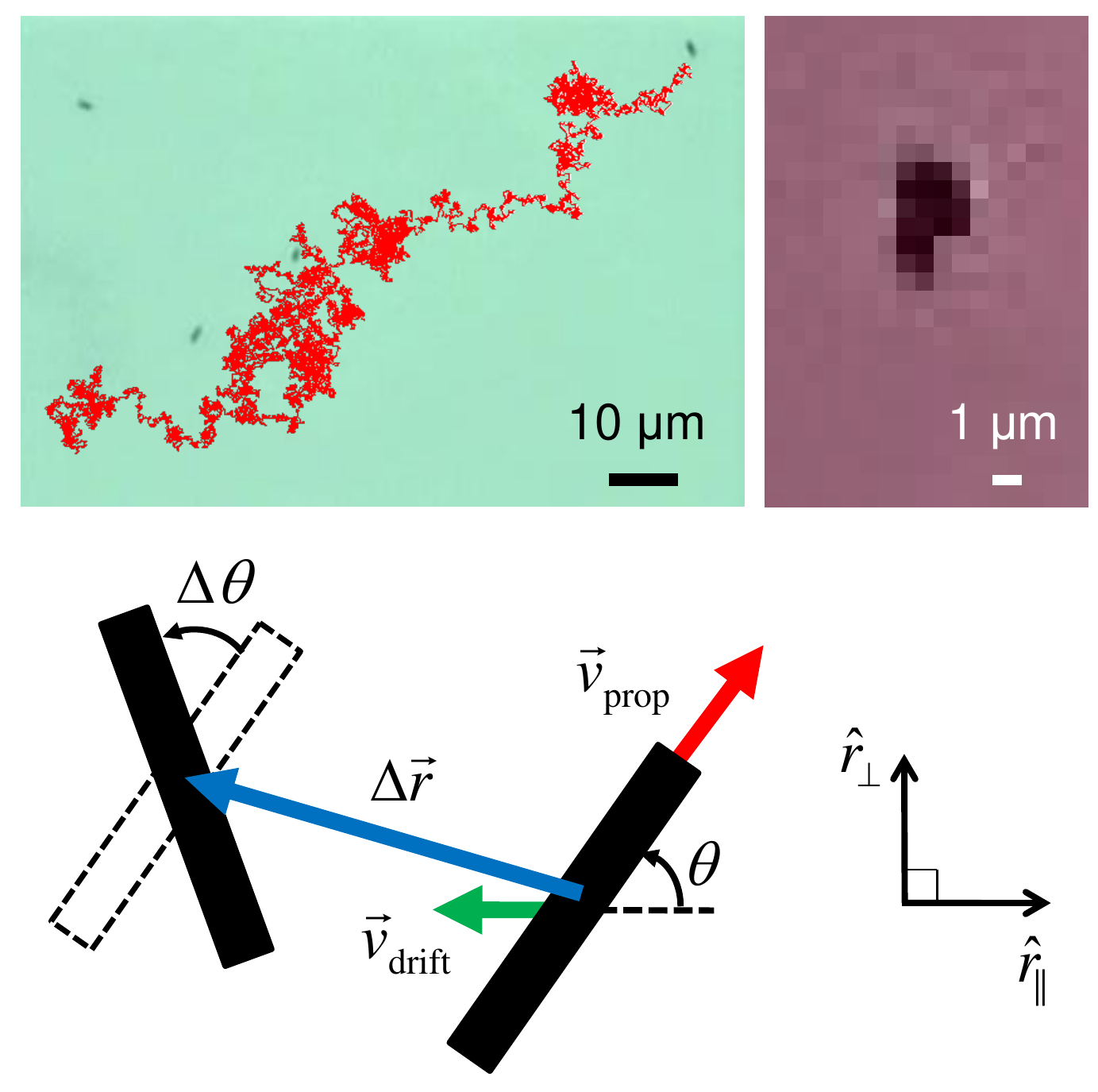}
\end{tabular}
\caption{(Color online) Top left: A video frame with a Pt/Au microrod with its trajectory superimposed, for microrod {\bf B} \textcolor{black}{(see text for the definition of microrod {\bf B})} (the hydrogel sits $\sim$0.5~mm off the left side of the frame). Top right: \textcolor{black}{A sample} image of an aggregate of multiple Pt/Au microrods from the video of Ref.~\cite{hong07}. Bottom: A representation of the frame-to-frame motion of the microrod. The $r_{\parallel}$ and $r_{\perp}$ axes (fixed in the lab frame) are antiparallel and perpendicular to \textcolor{black}{the direction of drift motion}, respectively. $\theta$ represents the angle of the propelling end of the swimmer (here, a Pt/Au microrod) from the positive $r_{\parallel}$ axis ($0^{\circ}\leq\theta<360^{\circ}$). $\Delta \vec{r}$ and $\Delta\theta$ represent the swimmer's frame-to-frame \textcolor{black}{translational} and angular displacements, respectively. $\vec{v}_{\text{prop}}$ and $\vec{v}_{\text{drift}}$ represent the swimmer's self-propulsion and drift velocities, respectively.}
\label{fig:schematic}
\end{figure}

If the {\it self-propelled} motion of a swimmer of axial symmetry causes it to move towards a fuel source, then its directed movement towards that source should correlate to the orientation of its propulsive axis. The correlation could take one of several forms; the two most straightforward are (I) a bias in the distribution of orientations $\theta$ for the symmetry axis that favors those which point the propelling end towards the source (i.e. $\theta = 180^{\circ}$ as defined in Fig.~\ref{fig:schematic}) or (II) an orientation-speed correlation wherein the swimmer propels itself more quickly when its propelling end is pointed towards the source. In other words, the swimmer must either spend more time with its propelling end pointed towards the source or move faster during the time when it is so pointed. Any net motion towards or away from the source that is independent of $\theta$ is likely unrelated to the self-propelled motion of a swimmer whose mechanism of motility is correlated to its structural axis. Note that the correlations in question may be weak and dominated by Brownian motion over short times, since chemotactic motions or \textcolor{black}{advective} drifts may need to accumulate over long times to become visible experimentally.

%a putative chemical gradient

To demonstrate the methodology to distinguish these two behaviors in noisy experimental data, we analyze a relatively long (\textcolor{black}{$\sim$5,000 seconds}), \textcolor{black}{high-temporal-resolution} (30 frames per second), high-\textcolor{black}{spatial-}resolution ($\sim$0.3~$\mu$m/pixel) video of Pt/Au microrods moving in an apparatus similar to that of Ref.~\cite{hong07}, available in Supplemental Material~\cite{supplemental}. Qualitatively, \textcolor{black}{all microrods (including all microrod aggregates)} move towards a hydrogel at an average drift speed of $\sim$0.1~$\mu$m/s. \textcolor{black}{(Note that this video was taken $\sim$10 minutes after a H$_2$O$_2$-soaked hydrogel was immersed into water. Note also that the drift speed varies in time and position. The origin of the variation will be discussed later.)} Continuous trajectories of the microrod's \textcolor{black}{geometric center} position and in-plane orientation were extracted from the video through automated image analysis \textcolor{black}{(a similar work is reported in Ref.~\cite{Dunderdale12})}; one such trajectory is superimposed on a sample video frame in Fig.~\ref{fig:schematic}. The Pt and Au ends of Pt/Au microrods are challenging to distinguish via optical microscopy of individual still-frame images and the microrods' self-propelled motion at this low H$_2$O$_2$ concentration is too weak to be discerned from the difference in position between successive frames. However, the propelling end of a given swimmer can still be unambiguously determined by tracking both ends over long times and identifying a consistent time-averaged bias in the motion along the rod axis towards one end of the swimmer. The video under analysis contains about ten Pt/Au microrods, three of which are suitable for detailed analysis. These three are denoted {\bf A}, {\bf B}, and {\bf C}, \textcolor{black}{which come into the view of the microscope in order at an interval of $\sim$20 minutes as shown in Table~\ref{tab:speeds}}. The other microrods either stay in the field of view too briefly to provide adequate data or else are so short that they \textcolor{black}{tumble} (i.e. rotate in a plane not parallel to the underlying glass slide) often enough that their ends cannot be tracked continuously. \textcolor{black}{(Note that the tumbling of microparticles occurs frequently in the flowing solution, while barely in the stationary one. For example, Pt/Au microrods {\bf A}, {\bf B}, and {\bf C} in the flowing H$_2$O$_2$ solution tumbled approximately 60, 120, 35 times per 10 minutes, respectively, but Au microrods in the stationary water did not at all.)}
% We eliminate the tumbling ones from the analysis, so we can drop the sentence below. (vhc)
% Note that the tumbling of microrods occurs frequently in the flowing solution, while barely in the stationary one. 
%high-frame-rate
%with {\bf A} being the closet to the hydrogel initially

\begin{figure}
\begin{tabular}{c}
{\includegraphics[width=0.40\textwidth]{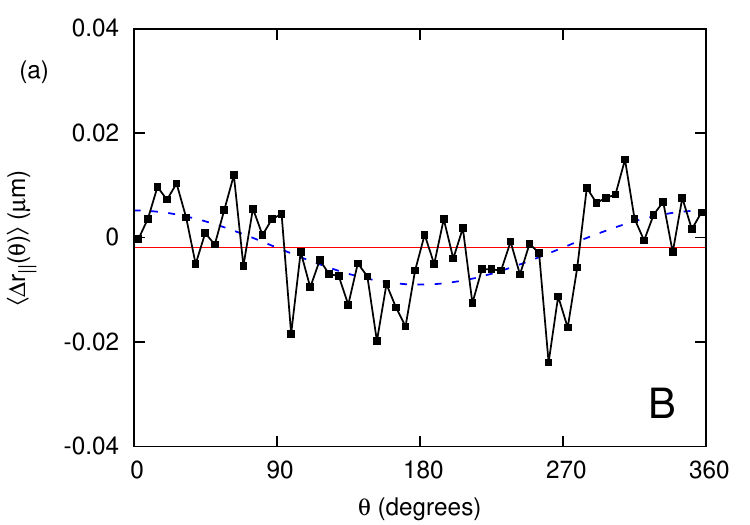}} \\
{\includegraphics[width=0.40\textwidth]{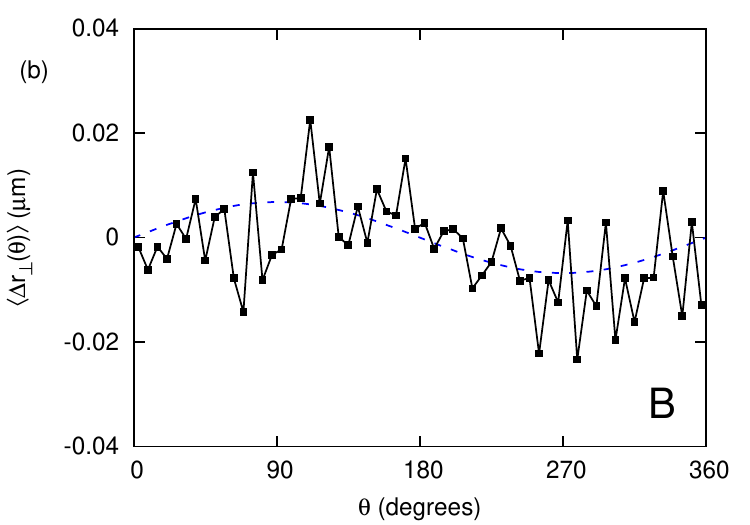}}
\end{tabular}
\caption{(Color online) (a) $\langle \Delta r_{\parallel} \rangle$ and (b) $\langle \Delta r_{\perp} \rangle$ for swimmer {\bf B} as a function of $\theta$ with a bin width of 6$^{\circ}$, time-averaged over approximately 400 samples for each $\theta$. The blue dashed lines are fitted to the data using the sinusoids of Eqs.~(\ref{Eq:assumed.delta.x}) and (\ref{Eq:assumed.delta.y}). \textcolor{black}{The red solid line shows the vertical shift of the sinusoid.} Swimmers {\bf A} and {\bf C} yield similar results \textcolor{black}{(see Supplemental Material~\cite{supplemental})}.}
\label{fig:dx.vs.theta}
\end{figure}
% Can we put the data for A and C into supplemental information, and note that fact here? (vhc)

The translational and rotational motions of sufficiently long Pt/Au microrods are generally in a plane parallel to the underlying glass slide (as is the case for any swimmer of axial symmetry that has a higher density than the ambient solution). Its state can thus be quantified by the spatial position ($\vec{r}$) of its \textcolor{black}{geometric center} and the angular orientation ($\theta$) of its propelling end, and its translational and rotational motions can be described by frame-to-frame \textcolor{black}{translational} and angular displacements ($\Delta \vec{r}$ and $\Delta \theta$, respectively) as shown in Fig.~\ref{fig:schematic}. $\Delta \vec{r}$ is decomposed into components $\Delta r_{\parallel}$ and $\Delta r_{\perp}$ \textcolor{black}{antiparallel and perpendicular to the direction of drift motion}. Figure~\ref{fig:dx.vs.theta} shows the microrod {\bf B}'s $\Delta r_{\parallel}$ and $\Delta r_{\perp}$ as a function of $\theta$, averaged over $\sim$400 observations in each 6-degree-wide angular bin. A sinusoidal pattern is clearly seen.

%parallel and perpendicular to a given direction of interest, for example the putative gradient in which it is immersed.
%center of mass (really, visual center)

We model swimmer motion, represented as $\Delta \vec{r}$, as the sum of three independent components: (i) self-propelled motion along the microrod's long axis towards the propelling end at speed $v_{\text{prop}}$, (ii) drift motion towards the hydrogel at speed $v_{\text{drift}}$ (of unspecified origin), and (iii) \textcolor{black}{translational} Brownian motion of the \textcolor{black}{microrod's geometric center}, with different diffusivities along and transverse to the microrod's long axis. \textcolor{black}{(Note that the microrod's net motion has one more component, rotational Brownian motion, but we do not take it into account in our investigation of the origin of chemotaxis because it has no effect on drift motion of a microrod. Instead, we present our analysis result for rotational diffusion of microswimmers {\bf A}, {\bf B}, and {\bf C} in Supplemental Material~\cite{supplemental}. Rotating microparticles yield similar results~\cite{Wang09,Amir13}.)} For the moment, we assume that $v_{\text{prop}}$ and $v_{\text{drift}}$ are independent of $\theta$, an assumption that will be confirmed by the analysis given below. In this model, $\Delta \vec{r}$ at a given $\theta$ follows the two-dimensional elliptical Gaussian probability density function, shifted from the origin by the self-propelled and drift terms:
\begin{align}
f(\Delta r_{\parallel}, \Delta r_{\perp}; \Delta t, \theta) &= N e^{-[a(\Delta r_{\parallel}')^2+b\Delta r_{\parallel}'\Delta r_{\perp}'+c(\Delta r_{\perp}')^2]}, \nonumber \\
\Delta r_{\parallel}' &= \Delta r_{\parallel} - (v_{\text{prop}}\cos\theta)\Delta t + v_{\text{drift}}\Delta t, \nonumber \\ 
\Delta r_{\perp}' &= \Delta r_{\perp} - (v_{\text{prop}}\sin\theta)\Delta t, \label{eq:gaussian.pdf}
\end{align}
where constants $N, a, b, c$ are given by
\begin{align}
N &= \frac{1}{4\pi\sqrt{D_{l}D_{s}}\Delta t}, \nonumber \\
a &= \frac{1}{4\Delta t}\left(\frac{\cos^{2}\theta}{D_{l}}+\frac{\sin^{2}\theta}{D_{s}}\right), \nonumber \\
b &= \frac{1}{4\Delta t}\left(\frac{-\sin2\theta}{D_{l}}+\frac{\sin2\theta}{D_{s}}\right), \nonumber \\
c &= \frac{1}{4\Delta t}\left(\frac{\sin^{2}\theta}{D_{l}}+\frac{\cos^{2}\theta}{D_{s}}\right), \nonumber
\end{align}
where $\Delta t$ is the time interval between successive frames and $D_{l}$, $D_{s}$ are the swimmer's diffusion coefficients along the long and short axes, respectively. Time-averaging Eq.~(\ref{eq:gaussian.pdf}) at a fixed $\theta$ yields the simple relations:
\begin{align}
\langle \Delta r_{\parallel}(\theta) \rangle &= (v_{\text{prop}} \cos\theta)\Delta t  - v_{\text{drift}}\Delta t, \label{Eq:assumed.delta.x} \\
\langle \Delta r_{\perp}(\theta) \rangle &= (v_{\text{prop}} \sin\theta ) \Delta t. \label{Eq:assumed.delta.y}
\end{align}

\begin{table}[t]
	\caption{The fitted values of self-propulsion speed $v_{\text{prop}}$ and drift speed $v_{\text{drift}}$ and the entrance times to the view of the video $t_{\text{in}}$ for three Pt/Au microrods.}
	\begin{tabular*}{0.48\textwidth}{@{\extracolsep{\fill}} c c c c}
		\hline \hline
		Swimmer & $v_{\text{prop}}$ ($\mu$m/s) & $v_{\text{drift}}$ ($\mu$m/s)& $t_{\text{in}}$ (mm:ss) \\ \hline
		{\bf A} & 0.46 & 0.099 & 00:00 \\
		{\bf B} & 0.21 & 0.057 & 26:25 \\
		{\bf C} & 0.15 & 0.093 & 39:48 \\
		\hline \hline
	\end{tabular*}	
	\label{tab:speeds}
\end{table}

The data of Figs.~\ref{fig:dx.vs.theta}(a), (b) fit this form well with self-propelled and drift speeds of constant magnitude, i.e. independent of $\theta$. The fitted values of $v_{\mathrm{prop}}$ and $v_{\mathrm{drift}}$ are listed in Table~\ref{tab:speeds} for the three swimmers examined. We find that $v_{\text{prop}}$ gradually decreases with time because Pt/Au microrods decompose H$_{2}$O$_{2}$ (i.e. consume the fuel) and therefore decrease its concentration. \textcolor{black}{[Note that the values of $v_{\mathrm{prop}}$ and $v_{\mathrm{drift}}$ in Table~\ref{tab:speeds} are temporally and spatially averaged. For example, $v_{\text{prop}}$ of a microswimmer coming into the view of the microscope (i.e. on the right of the video) is greater than that going out of the view of the microscope (i.e. on the left of the video). We did not take the temporal and spatial variation of the fuel concentration into account in our analysis because (i) a microswimmer spends $\sim$600 seconds in the view of the microscope, and for the $\sim$600 seconds, its $v_{\text{prop}}$ decreases only by $\sim$30\% as shown in Table~\ref{tab:speeds}, and (ii) if we do that, the statistical noise increases due to the insufficient number of frames. The temporal and spatial variation of $v_{\mathrm{drift}}$ will be discussed later.]} Some residual fluctuations survive the time averaging; these appear to be purely Brownian in origin: non-self-propelled Au microrods with the same size and shape as Pt/Au ones in water show similar residuals, and the root-mean-square deviations from the sinusoidal fits are consistent with the values of $D_{l}$ and $D_{s}$ extracted from the same dataset. \textcolor{black}{[Note that the statistical noise is comparable to the amplitude of the sinusoids, which is proportional to $v_{\text{prop}}$, because $v_{\text{prop}}$ is very small ($\sim$0.2~$\mu$m/s) due to the very low H$_{2}$O$_{2}$ concentration. The large $v_{\text{prop}}$ at the high H$_{2}$O$_{2}$ concentration (e.g. 6.6~$\mu$m/s at 3.3\% in Ref.~\cite{paxton04}) makes the statistical noise negligible compared to the amplitude of the sinusoids.]} Also, the drift speeds obtained by simply dividing the total \textcolor{black}{translational} displacements by the elapsed time (0.12, 0.075, and 0.099 $\mu$m/s, respectively, for microrods {\bf A}, {\bf B}, and {\bf C}) are consistent with the fitted ones (0.099, 0.057, and 0.093 $\mu$m/s).

\begin{figure}
%\begin{center}
%\includegraphics[trim=0mm 0mm 40mm 0mm, clip, width=0.40\textwidth]{figure_multiplot}
\includegraphics[trim=10mm 5mm 45mm 15mm, clip, width=0.40\textwidth]{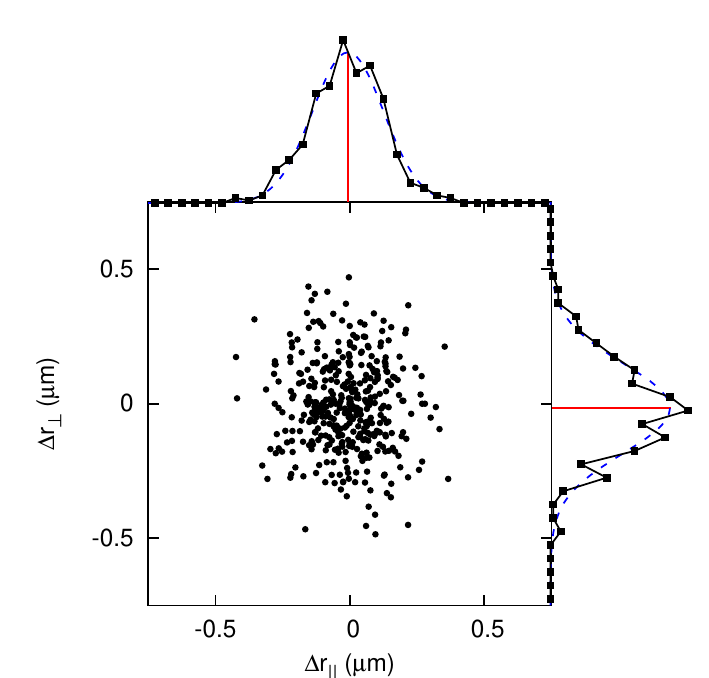}
%\end{center}
\caption{(Color online) Scatterplot of $\Delta \vec{r}$ for the Pt/Au microrod {\bf B} at $267^{\circ}<\theta<273^{\circ}$. Each point represents a single frame-to-frame \textcolor{black}{translational} displacement of the microrod, 396 total. The blue dashed lines are fitted to the projections of the scatterplot onto $r_{\parallel}$ and $r_{\perp}$ axes with a bin width of 0.05~$\mu$m using the Gaussian function of Eq.~(\ref{eq:gaussian.pdf}) with $\theta=270^{\circ}$. \textcolor{black}{The red solid lines show the horizontal and vertical shifts of the Gaussian function.}}
\label{fig:gaussian.function}
\end{figure}

To further verify the description of Eq.~(\ref{eq:gaussian.pdf}), all observed values of $\Delta \vec{r}$ for the Pt/Au microrod {\bf B} at $267^{\circ}<\theta<273^{\circ}$ are depicted in Fig.~\ref{fig:gaussian.function}. (Note that at $\theta=270^{\circ}$, self-propelled and drift motions are orthogonal and thus easier to distinguish.) The distribution of $\Delta \vec{r}$ is in excellent agreement with Eq.~(\ref{eq:gaussian.pdf}), as evidenced by normalized distributions of $\Delta r_{\parallel}$ and $\Delta r_{\perp}$, yielding $D_{s}= 0.26$ $\mu$m$^{2}$/s and $D_{l} = 0.42$ $\mu$m$^{2}$/s. These coefficients are generally independent of $\theta$ (e.g. $\theta=90^{\circ}$ yields $D_{s}= 0.27$ $\mu$m$^{2}$/s and $D_{l} = 0.42$ $\mu$m$^{2}$/s). The center of the Gaussian distribution is shifted slightly towards the hydrogel (left) due to drift and towards the propelling end (down) due to self-propulsion, in both cases by amounts that are consistent with $v_{\text{prop}}\Delta t$ and $v_{\text{drift}}\Delta t$. \textcolor{black}{[Note that the two shifts of the Gaussian distribution are not clearly visible because at $\Delta t = 1/30$ second, displacements by self-propelled and drift motions are much smaller than that by translational Brownian motion (i.e. by $\sim$50 times).]} The ratio $D_{l}/D_{s} \approx 1.6$ is the same as that observed for the Au microrods in water (0.20 vs. 0.32 $\mu$m$^{2}$/s for one Au microrod, 0.14 vs. 0.23 $\mu$m$^{2}$/s for another). [Note that $D_{l}$ and $D_{s}$ for Au microrods were obtained from $\sim$10,000 \textcolor{black}{translational displacements measured in} the instantaneous rest frame of the microrod owing to the absence of drift motion \textcolor{black}{(see Supplemental Material~\cite{supplemental})}. \textcolor{black}{Note also that the two Au microrods have similar shapes and sizes, but their translational diffusion coefficeints are different by $\sim$30\%.}]

\begin{figure}
\begin{tabular}{c}
%{\includegraphics[width=0.40\textwidth]{figure_dx_dy_dr_vs_dtheta_pt_au_gel}} \\
{\includegraphics[width=0.40\textwidth]{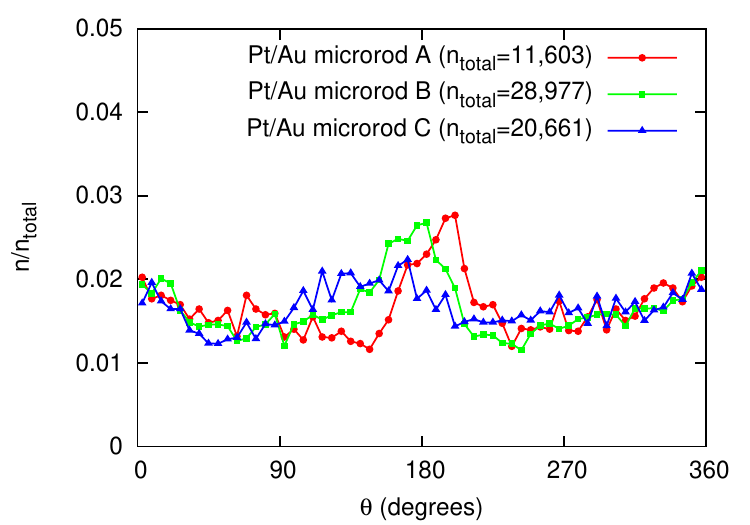}}
\end{tabular}
\caption{(Color online) Normalized distributions of $\theta$ for Pt/Au microrods {\bf A}, {\bf B}, and {\bf C} with a bin width of 6$^{\circ}$. $n$ and $n_{\mathrm{total}}$ represent the number of frames at $\theta$ and the total number of frames analyzed, respectively.}
\label{fig:orientation}
\end{figure}

These results can now be analyzed to determine the applicability of mechanisms (I) and (II). The simple sinusoidal variation of $\vec{v}_{\text{prop}}$ with respect to $\theta$ indicates no correlation between the microrod's self-propulsion speed and orientation; hence, mechanism (II) is not the case. To see if mechanism (I) applies here, we plot normalized distributions of $\theta$ for swimmers {\bf A}, {\bf B}, and {\bf C} in \textcolor{black}{Fig.~\ref{fig:orientation}}. The distributions show a slight excess of orientations pointing towards the hydrogel ($90^{\circ} < \theta < 270^{\circ}$): 49.8\%, 51.5\%, and 52.4\%, respectively. \textcolor{black}{(Note that our video's low spatial resolution and non-square pixel aspect ratio hinder the accurate determination of $\theta$.)} Using the fitted $v_{\text{prop}}$, the consequences of this asymmetry in terms of biased drift can be straightforwardly computed by angular averaging over this distribution: the resulting average drift speed can account for only a small fraction of the observed bias of the trajectory (2.9\%, 10.1\%, and 4.5\% for {\bf A}, {\bf B}, and {\bf C}, respectively). Therefore, mechanism (I) is not the primary source of the observed drift in this system. Thus, large-scale full-trajectory analysis can provide unambiguous conclusions about subtle sources of mobility, even when the individual frame-by-frame motions of swimmers are dominated by Brownian fluctations. Further \textcolor{black}{quantitative} information on real gradients and flow fields that is potentially available in controlled microfluidic geometries such as those afforded by microchannels~\cite{jeon02} could also be naturally incorporated into such an analysis.

% for the videos in question

In cases where long-term trajectories of multiple swimmers are available, it is also informative to look for correlations between $v_{\text{prop}}$ and $v_{\text{drift}}$ across the observed swimmer population, although the statistical power of such an analysis will be limited by the number of available swimmers with sufficient trajectory information. In the current case, the results of Table~\ref{tab:speeds} show for example that swimmers {\bf A} and {\bf C} have very similar $v_{\text{drift}}$ (0.099 and 0.093 $\mu$m/s), but very different $v_{\text{prop}}$ (0.46 and 0.15 $\mu$m/s). Since chemotactic drift derives from powered motion, a lack of correlation between the magnitudes of drift and propulsion across a swimmer population would argue towards a passive \textcolor{black}{advective} mechanism and conversely, a positive correlation between these two quantities would argue for a powered chemotactic origin to any observed long-time, long-distance drift. \textcolor{black}{[Note that the three microswimmers have different values of $v_{\text{drift}}$ because the bulk flow generated by a hydrogel is unsteady and non-uniform (i.e. varies in time and position). As a hydrogel absorbs water, its water-absorbing capacity and speed decrease with time~\cite{Kabiri11, Ahmed15} and thus the speed of the bulk flow decreases with time. This explains why microparticles' $v_{\text{drift}}$ decreases with time in Ref.~\cite{hong07}. Also, the speed of the bulk flow created by a hydrogel depends on position (i.e. microparticles accumulate at a certain location of a hydrogel in Ref.~\cite{hong07}) and distance (i.e. microparticles closer to a hydrogel drift faster in Ref.~\cite{hong07}).]} Similar reasoning applies to swimmers whose powered motility may be impeded by attachment of cargo~\cite{sun08}, those whose long axis is sufficiently short that it spends significant time outside the plane of the slide, or more geometrically complex swimmer aggregates (as shown in the top right panel of Fig.~\ref{fig:schematic}) that have a different balance of translational and rotational motion \textcolor{black}{(e.g. rotating motion, orbiting motion, or no self-propelled motion)}~\cite{Mirkovic10b}. Similarly, if control experiments with unpowered particles or \textcolor{black}{non-fuel sources} yield similar drift motions to those observed for powered swimmers \textcolor{black}{near fuel sources} (as is the case for a reproduction of \textcolor{black}{old hydrogel and capillary} experiments in Ref.~\cite{hong07}), bulk \textcolor{black}{advective} flows (such as could be induced by the accommodation of a H$_2$O$_2$-soaked hydrogel to a deionized water environment) are the most plausible source of drift. \textcolor{black}{(Note that only three measurements may be too small to validate our statistical data analysis method. However, instead of performing the same experiment again to obtain a larger number of measurements, we chose to verify the prediction from the three measurements and reveal the origin of chemotaxis by performing other experiments in Ref.~\cite{hong07}. Note also that we already applied our analysis method to rotating microparticles and revealed the effect of their geometry on their function~\cite{Amir13}, which is another example showing the capability of our method.)}

%non-fuel gradients
%in fuel gradients
%the controls in prior \textcolor{black}{hydrogel and capillary} experiments in Ref.~\cite{hong07}

\textcolor{black}{In conclusion, we developed novel data analysis methods to investigate the microscopic origins of the macroscopic collective behaviors of self-propelled microparticles: (i) the examination of microparticle aggregates, which gives a clue to how the size and shape of microparticles and the type and strength of self-propelled motion affect microparticles' collective behaviors \textcolor{black}{(this method is qualitative, but efficient)}, and (ii) the statistical analysis of translational and rotational trajectories of microparticles, which directly shows how the microscopic self-propelled motion and the macroscopic collective motion \textcolor{black}{(e.g. drift motion)} are related to each other \textcolor{black}{(this method is quantitative and accurate)}. As an example, we applied our analysis methods to the directed motion of \textcolor{black}{catalytically-driven} microparticles reported in Refs.~\cite{hong07, sun08, hong10} and revealed the origin of the decrease in the migration speed of microparticles with time and distance, \textcolor{black}{which was verified by new experiments}. \textcolor{black}{This suggests that our methods can be useful tools} to uncover the unknown origins of other exotic collective motions of microswimmers.}

\begin{acknowledgments}
We are grateful to Eric Mockensturm for helpful discussions \textcolor{black}{and to Shakuntala Sundararajan for the valuable information}. We acknowledge the support of NSF DMR-1420620.
\end{acknowledgments}

\bibliography{Chemotaxis}

\end{document}